\documentclass[12pt]{article}
\usepackage{epsf}

\def\lesssim{{\
\lower-1.2pt\vbox{\hbox{\rlap{$<$}\lower5pt\vbox{\hbox{$\sim$}}}}\ }} 
\def\gtrsim{{\
\lower-1.2pt\vbox{\hbox{\rlap{$>$}\lower5pt\vbox{\hbox{$\sim$}}}}\ }}

\begin{document}

\begin{titlepage}

\begin{flushright}
    January 1997
\end{flushright}

\vspace{0.5cm}

\begin{center}
    {\Large\bf Constraint on Cosmic Density of the String
    Moduli Field in Gauge-Mediated Supersymmetry-Breaking Theories}\\
    \vspace{1.5cm} 
    {\large M.~Kawasaki$^{a,*}$ and
    T.~Yanagida$^{b,\dagger}$}\\
    \vspace{1cm}
    {\it $^{a}$Institute for Cosmic Ray Research, University of Tokyo,
    Tokyo 188, Japan\\
    $^{b}$Department of Physics, University of Tokyo, Tokyo 133, Japan\\
    $^{*}$e-mail:kawasaki@icrr.u-tokyo.ac.jp\\
    $^{\dagger}$e-mail:yanagida@kanquro.phys.s.u-tokyo.ac.jp}
\end{center}

\vspace{2.0cm}

\begin{abstract}
    We derive a constraint on the cosmic density of string moduli
    fields in gauge-mediated supersymmetry-breaking theories by
    requiring that photons emitted from the unstable moduli fields
    should not exceed the observed X-ray backgrounds.  Since mass of
    the moduli field lies in the range between $O(0.1)$keV and
    $O(1)$MeV and the decay occurs through a gravitational
    interaction, the lifetime of the moduli field is much longer than
    the age of the present universe. The obtained upperbound on their
    cosmic density becomes more stringent than that from the unclosure
    condition for the present universe for the mass greater than about
    100keV.
\end{abstract}


\end{titlepage}
\newpage


\section{Introduction}

Massless moduli fields $\phi$ exist in all known superstring theories
which parameterize continuous vacuum-state
degeneracies~\cite{GSW}. They are expected to get their masses from
nonperturbative dynamics which breaks the supersymmetry (SUSY) and a
generic argument~\cite{Carlos} shows that their masses are comparable
to the gravitino mass $m_{3/2}$. In hidden sector models of SUSY
breaking the moduli fields have the masses at the electroweak
scale. On the other hand, their masses lie in the keV range in
gauge-mediated SUSY-breaking
models~\cite{Dine,Intriligator,Gouvea}. In the latter models the
lifetimes of the moduli fields are much longer than the age of the
present universe and energy densities of their coherent oscillations
easily exceed the critical density of the universe. Thus the
gauge-mediated SUSY-breaking models looks to contradict the
superstring theories as long as the light moduli fields exist in the
keV region.\footnote{
If some string dynamics gives rise to large SUSY-invariant masses for
the moduli fields, there is no contradiction between gauge-mediated
SUSY-breaking and superstring theories. However, no compelling model
has been found so far.}

In a recent paper~\cite{Gouvea}, Gouv\^ea, Moroi and Murayama have
pointed out that the above moduli problem may be solved if a late-time 
thermal inflation takes place. In this letter we show that a stringent 
constraint on the cosmic density of the moduli fields is derived from
the experimental upperbounds on the cosmic X-ray backgrounds. The
basic assumption in the present analysis is that the main decay mode
of the moduli field $\phi$ is a two-photon process, $\phi \rightarrow
2\gamma$. This assumption is quite reasonable since the decay mode to
two neutrinos, $\phi \rightarrow \nu_{L} + \bar{\nu}_{L}$, has a
chirality suppression and vanishes for massless neutrinos.\footnote{
The process, $\phi \rightarrow \nu_{L} + \nu_{L}$, breaks the
electroweak gauge symmetry and hence it has an extra suppression.}

\section{Constraint from X-ray Background}

Here we perform a model-independent analysis taking the mass
$m_{\phi}$, the lifetime $\tau_{\phi}$ and the energy density
$\rho_{\phi}$ of the moduli field as free parameters. We consider the
mass region between $\sim$ 0.2keV and  $\sim$ 4MeV.

First we summarize the observational fluxes $F_{\gamma,obs}$ of the
cosmic X-ray backgrounds in the corresponding photon energy region
$E_{\gamma} \simeq 0.1\textrm{keV} \sim 2\textrm{MeV}$. The ASCA
satellite experiment measures the flux $F_{\gamma}$ for
$0.1\textrm{keV} \le E_{\gamma} \le 7\textrm{keV}$~\cite{Gendreau}.
For higher photon energies, HEAO satellite gives useful
data~\cite{HEAO}. We can fit these observational data by simple three
power-low spectra for $0.1\textrm{keV} \lesssim E_{\gamma} \lesssim
1\textrm{MeV}$:
\begin{eqnarray}
    F_{\gamma, obs}(E_{\gamma}) & \simeq  & 
    8 (E_{\gamma}/\textrm{keV})^{-0.4}
    ~~~~~~~ 0.1\textrm{keV} \lesssim E_{\gamma} \lesssim 25\textrm{keV},
    \label{obs1}\\
     & \simeq  & 380 (E_{\gamma}/\textrm{keV})^{-1.6}
    ~~~~~ 25\textrm{keV} \lesssim  E_{\gamma} \lesssim 350\textrm{keV},
    \label{obs2}\\
     & \simeq  & 2 (E_{\gamma}/\textrm{keV})^{-0.7}
    ~~~~~~~ 350\textrm{keV} \lesssim  E_{\gamma} \lesssim 2\textrm{MeV},
    \label{obs3}
\end{eqnarray}
where $F_{\gamma, obs}$ is measured in units of $(\textrm{cm$^2$ sr
sec})^{-1}$. (The fit is not good for $E_{\gamma} \gtrsim 1$MeV since 
data scatter very much.)

Next let us  estimate the photon flux from decaying moduli field. The
present cosmic number density $n_{\phi}$ of the moduli field is given
by its mass $m_{\phi}$ and density $\rho_{\phi}$ as 
\begin{equation}
    \label{num-density}
    n_{\phi} = 10.54 \textrm{cm}^{-3} (m_{\phi}/\textrm{keV})^{-1} 
    (\Omega_{\phi}h^2),
\end{equation}
where $\Omega_{\phi} \equiv \rho_{\phi}/\rho_{c}$ ($\rho_{c}$:
critical density of the universe) and $h$ is the present Hubble
constant in units of 100km/sec/Mpc. Since two monochromatic photons
with energy $m_{\phi}/2$ are produced in the decay, the flux from the
moduli fields is estimated as
\begin{eqnarray}
    F_{\gamma}(E_{\gamma}) & = & \frac{E_{\gamma}}{4\pi} \int_{0}^{t_0} dt'
    \frac{1}{\tau_{\phi}} n_{\phi}
    (1+z) 2 \delta (E_{\gamma}(1+z) -m_{\phi}/2),\nonumber \\
     & = & \frac{\sqrt{2}n_{\phi}E_{\gamma}^{3/2}}
     {\pi\tau_{\phi}H_0m_{\phi}^{3/2}}
     [ \Omega_0 + (1-\Omega_0-\Omega_{\Lambda})(2E_{\gamma}/m_{\phi})
     + \Omega_{\Lambda})(2E_{\gamma}/m_{\phi})^{3}]^{-1/2} , 
     \label{flux}
\end{eqnarray}
where $t_0$ is the present time, $z$ is the redshift, $H_0$ is the
present Hubble constant, $\Omega_0$ is the present (total) density
parameter and $\Omega_{\Lambda}$ is the density parameter of the
cosmological constant. The flux $F_{\gamma}$ takes a maximum value
$F_{\gamma,max}$ at $E_{\gamma} = m_{\phi}/2$. From
eqs.(\ref{num-density}) and (\ref{flux}) $F_{\gamma,max}$ is given by
\begin{equation}
    \label{max-flux}
    F_{\gamma,max} = 1.55\times 10^3 (\textrm{cm$^2$ str sec})^{-1}
    \left(\frac{m_{\phi}}{\textrm{keV}}\right)^{-1}
    \left(\frac{\tau_{\phi}}{10^{25}\textrm{sec}}\right)^{-1}
    (\Omega_{\phi}h). 
\end{equation}

By requiring that $F_{\gamma,max}$ should be less than the observed
X-ray background fluxes $F_{\gamma,obs}$ (eqs.(\ref{obs1}) --
(\ref{obs3})), we can obtain the constraint on $\Omega_{\phi}$ using
eqs.(\ref{obs1})--(\ref{obs3}):
\begin{eqnarray}
    \Omega_{\phi}h  
    & \lesssim & 7\times 10^{-3} 
    \left (\frac{\tau_{\phi}}{10^{25\sec}}\right)
    \left(\frac{m_{\phi}}{\textrm{keV}}\right)^{0.6} 
    ~~~~~~ 0.2\textrm{keV} \lesssim m_{\phi} \lesssim 50\textrm{keV},\\
    & \lesssim & 7\times 10^{-1} 
    \left (\frac{\tau_{\phi}}{10^{25\sec}}\right)
    \left(\frac{m_{\phi}}{\textrm{keV}}\right)^{-0.6} 
    ~~~~~ 50\textrm{keV} \lesssim   m_{\phi} \lesssim 0.7\textrm{MeV},\\
    & \lesssim & 2\times 10^{-3} 
    \left (\frac{\tau_{\phi}}{10^{25\sec}}\right)
    \left(\frac{m_{\phi}}{\textrm{keV}}\right)^{0.3}
    ~~~~~~ 0.7\textrm{MeV} \lesssim  m_{\phi} \lesssim  2\textrm{MeV}.
\end{eqnarray}
This constraint is shown in Fig.~1 by the dashed line.  The use of the
simple power-low spectra (\ref{obs1})--(\ref{obs3}) is convenient to
get the analytic expression for the upperbound on $\Omega_{\phi}$.
However, a more precise constraint is obtained by using a smooth curve
which gives better fit to observational data and the resultant
constraint on $\Omega_{\phi}/\tau_{\phi}$ is shown in Fig.~1 by solid
curve.

\section{Gravitational Decay of the String Moduli Field}
\label{sec:constraint}

The constraint obtained above is general in the sense that it applies
to any unstable particles which decay mainly into photons with
lifetime much longer than the age of the universe.  We now apply this
model-independent constraint to the string moduli field which decays
into photons only through a gravitational interaction. The lifetime of
the moduli
field $\tau_{\phi}$ is related to its mass $m_{\phi}$ by\footnote{
The decay rate $\Gamma_{\phi}$ of the moduli field has an unknown
parameter $h$ of order 1 as $\Gamma_{\phi}\simeq
hm_{\phi}^3/M_{p}^2$. If one uses smaller values of $h$, $h< 1$, the
obtained constraint becomes weaker.}
\begin{equation}
    \label{lifetime}
    \tau_{\phi} \simeq \frac{M_{p}^2}{m_{\phi}^3} \simeq 
    10^{32}\textrm{sec} \left(\frac{m_{\phi}}{\textrm{keV}}\right)^{-3},
\end{equation}
where $M_{p}$ is the Planck mass. With this relation, the maximum flux 
$F_{\gamma,max}$ is given by
\begin{equation}
    \label{max-flux2}
    F_{\gamma,max} = 1.55\times 10^{-4}(\textrm{cm$^2$ str sec})^{-1}
    \left(\frac{m_{\phi}}{\textrm{keV}}\right)^{2}
    (\Omega_{\phi}h). 
\end{equation}
Then, in the same way as in the previous section, we can obtain the
constraint on $\Omega_{\phi}$ which is shown in Fig.~2. From the
figure it is seen that the upperbound on $\Omega_{\phi}$ is less than
1 for $m_{\phi} \gtrsim 100$keV. Thus, for the moduli field with such
masses the constraint from the X-ray backgrounds is much more
stringent than that from the critical density (i.e. $\Omega_{\phi}
\lesssim 1$).\footnote{
Here we assume that the density of the moduli fields are homogeneously
in the universe. However, if the moduli fields concentrates in the
halo of our galaxy, one may obtain the more stringent constraint.  }

In order to satisfy the constraint on $\Omega_{\phi}$ it is necessary
to dilute the density of moduli fields by some large entropy production such
as thermal inflation~\cite{Lyth,Gouvea}.  However, the entropy
production cannot be too large since it also dilutes the baryon number
in the universe. For example, in ref.~\cite{Gouvea} it has been shown 
that the Affleck-Dine baryogenesis can generate the baryon asymmetry
given by
\begin{equation}
   \label{baryon}
   \frac{n_B}{s} \lesssim 4\times 10^{-5} (\Omega_{\phi}h^2)
   \left(\frac{m_{3/2}}{100\textrm{keV}}\right)^{-1},
\end{equation}
where $n_B$ is the baryon number density and $s$  the entropy 
density. Requiring $n_B/s \simeq 4\times 10^{-11}$, we obtain the
lower limit on $\Omega_{\phi}$:
\begin{equation}
    \label{baryon-const}
    \Omega_{\phi}h^2 \gtrsim 10^{-6}
    \left(\frac{m_{\phi}}{100\textrm{keV}}\right),
\end{equation}
where we assume $m_{\phi} \simeq m_{3/2}$. This, together 
with the X-ray background constraint, leads to $m_{\phi} \lesssim 
2$MeV (see Fig.~2). 

\section{Conclusion}

We have shown that a stringent constraint on the density of the string
moduli fields in gauge-mediated SUSY-breaking theories is obtained by
requiring that the photons emitted from the unstable moduli field
should not exceed the observed X-ray backgrounds.  In particular, if
the decay occurs through the gravitational interaction, the lifetime
of the moduli field is estimated as a function of its mass and the
constraint on $\Omega_{\phi}$ is more stringent than that from the
unclosure condition for the present universe for $m_{\phi} \gtrsim
100$keV.

The observed X-ray backgrounds are usually explained by the emission
from active galactic nuclei and Type I supernova.  However it is known
that there may be a hump~\cite{HEAO} in the spectrum at a few MeV
which can not be explained by usual sources. It is very much
intriguing that the unstable moduli fields with mass about 1MeV are
indeed the source of the hump. However the recent COMPTEL
observation~\cite{Kappadath} has not confirmed the existence of the
hump. Thus the precise measurement of the spectrum at $O(1)$MeV
energies by future experiments are highly required.

\newpage

\noindent\textbf{Figure Captions}

\begin{description}
\item[Fig.1] Upperbound on $\Omega_{\phi}/\tau_{\phi}$ from the X-ray 
background radiation (solid curve). The dashed line represents the 
constraint obtained by using the simple fitting spectrum of the 
X-ray background. 
\item[Fig.2] Upperbound on $\Omega_{\phi}$ from the X-ray background 
(solid curve). The lowerbound  from the baryogenesis (proposed in 
ref.\cite{Gouvea}) is denoted by the dashed line.
\end{description}

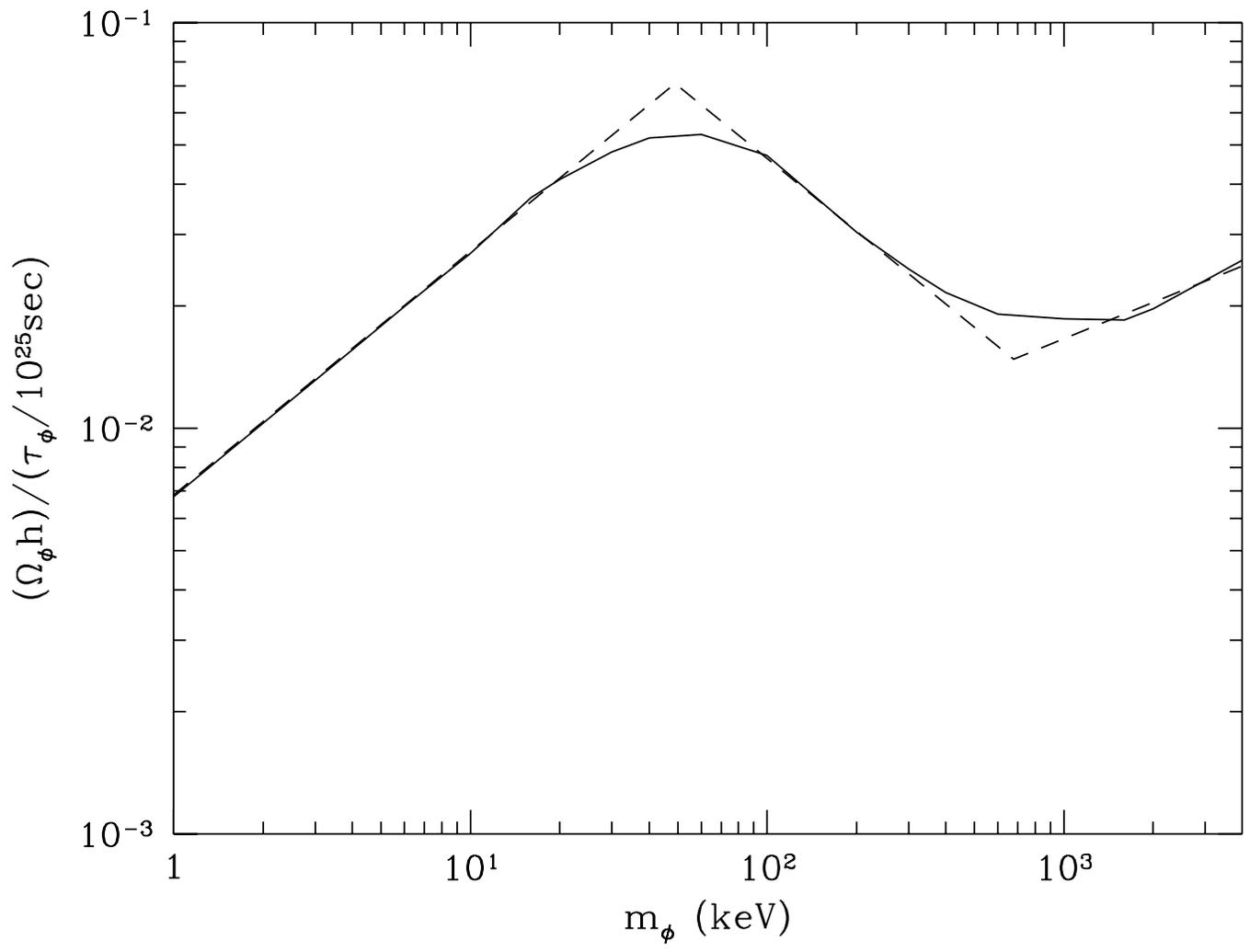
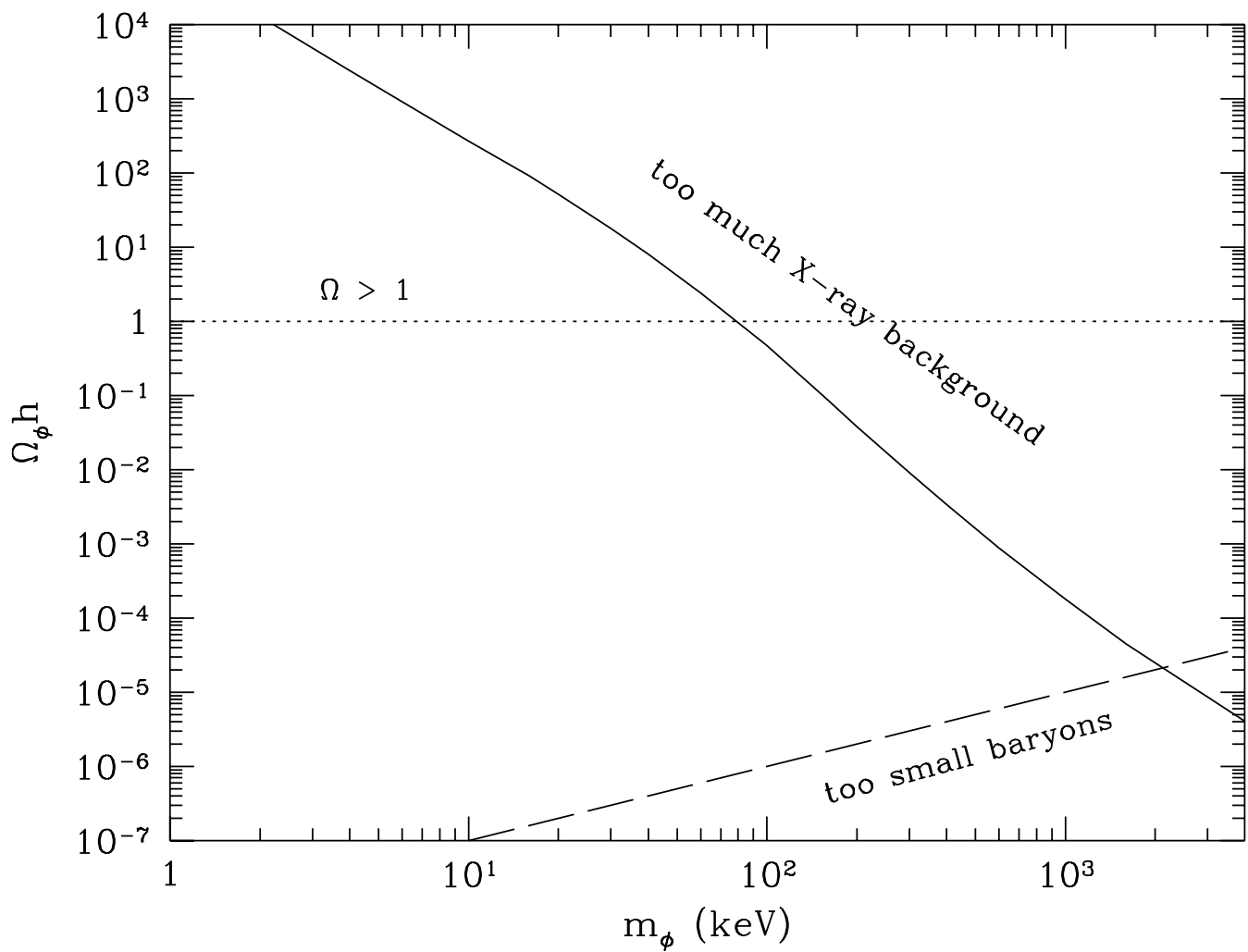
\end{document}